# Static Hopf Solitons and Knotted Emergent Fields in Solid-State Noncentrosymmetric Magnetic Nanostructures

Jung-Shen B. Tai (戴榮身)[1] and Ivan I. Smalyukh[1,2,3]*

[1]Department of Physics, University of Colorado, Boulder, Colorado 80309, USA
[2]Materials Science and Engineering Program, Soft Materials Research Center and Department of Electrical, Computer and Energy Engineering, University of Colorado, Boulder, Colorado 80309, USA
[3]Renewable and Sustainable Energy Institute, National Renewable Energy Laboratory and University of Colorado, Boulder, Colorado 80309, USA

*Corresponding author: ivan.smalyukh@colorado.edu

*Two-dimensional topological solitons, commonly called Skyrmions, are extensively studied in solid-state magnetic nanostructures and promise many spintronics applications. However, three-dimensional topological solitons dubbed hopfions have not been demonstrated as stable spatially localized structures in solid-state magnetic materials. Here we model the existence of such static solitons with different Hopf index values in noncentrosymmetric solid magnetic nanostructures with a perpendicular interfacial magnetic anisotropy. We show how this surface anisotropy, along with the Dzyaloshinskii-Moriya interactions and the geometry of nanostructures, stabilize hopfions. We demonstrate knots in emergent field lines and computer simulate Lorentz transmission electron microscopy images of such solitonic configurations to guide their experimental discovery in magnetic solids.*

Topological solitons exist in the effective field theories of many physical systems [1-9]. For example, two-dimensional (2D) particlelike Skyrmions are solitonic field configurations classified by elements of the second homotopy group $\pi_2(\mathbb{S}^2) = \mathbb{Z}$ and indexed by a topological charge, the Skyrmions number. Such Skyrmions are widely studied in solid magnets due to the wealth of new fundamental physical phenomena and potential for technological applications [9-13]. Of particular interest is the realization of solitons in various nanostructures, like thin films and channels, which may enable new breeds of magnetic memory units and spintronics applications [9–13]. Three-dimensional (3D) Hopf solitons, or hopfions, are field configurations localized in all three spatial dimensions, embedded in a uniform far field and identified maps from $\mathbb{R}^3 \cup \{\infty\} \cong \mathbb{S}^3$ to the order-parameter space (target space) $\mathbb{S}^2$ of three-dimensional unit vectors; they belong to the third homotopy group $\pi_3(\mathbb{S}^2) = \mathbb{Z}$. Topologically distinct hopfions are characterized by the Hopf index $Q \in \mathbb{Z}$ with a geometric interpretation of the linking number of any two closed-loop preimages [14], regions in space with the same orientation of field corresponding to a single point on $\mathbb{S}^2$. Hopfions were predicted in many physical systems [1-3,6-8,15-21] and stable static hopfions were recently demonstrated experimentally in liquid crystals [17] and chiral colloidal ferromagnets [18,19] through direct 3D imaging and numerical simulations. Dynamically propagating or precessing hopfions were modeled in ferromagnets [15,16,22] while static hopfions comprising knots of Skyrmions were considered in frustrated magnets [8], though host materials with the required frustration remain to be identified. The feasibility of realizing stable hopfions in widely studied chiral magnets or magnetic nanostructures remains unknown.

In this work, we perform numerical simulations and predict stable static hopfions in noncentrosymmetric magnetic nanostructures with perpendicular magnetic anisotropy (PMA) at their interfaces. We show that, in addition to Dzyaloshinskii-Moriya interactions (DMI) [10], confinement and interfacial PMA stabilize hopfions [23]. We focus on fully nonsingular field configurations, Skyrmions and hopfions with different Hopf indices, and study knots in preimages and in the emergent field associated with them; solitons accompanied by singular point defects (Bloch points), such as chiral bobbers [24] and torons [25], will be explored elsewhere. To facilitate experimental discovery of these structures, we construct diagrams of the structural stability of the localized field configurations versus material and geometric parameters and applied magnetic field. We also numerically simulate their Lorentz transmission electron microscopy (TEM) images and discuss how hopfions can be identified using real-space imaging techniques [11,26,27].

A computer simulated structure of magnetization field $\boldsymbol{m}(\boldsymbol{r})$ of the Hopf soliton is shown in Fig. 1. It features closed-loop preimages corresponding to all points of $\mathbb{S}^2$, with each pair of distinct preimages linked the same $Q$ number of times. The exterior of the torus-embedded region is occupied by the preimage of the point in $\mathbb{S}^2$ corresponding to the far-field background $\boldsymbol{m}_0$ (set along $+\hat{z}$), within which all other preimages are smoothly embedded (Fig. 1). Such elementary Hopf solitons comprise interlinked closed-loop preimages of constant $\boldsymbol{m}(\boldsymbol{r})$, in this resembling the topology of mathematical Hopf fibration [28]. Because of the field topology, the emergent magnetic field $(\boldsymbol{B}_{\text{em}})_i \equiv \hbar \varepsilon^{ijk} \boldsymbol{m} \cdot (\partial_j \boldsymbol{m} \times \partial_k \boldsymbol{m})/2$ [29-31] of a solid-state elementary hopfion spirals around its symmetry axis with a unit flux quantum [Figs. 1(c) and 1(d)]. Remarkably, each pair of streamlines of $\boldsymbol{B}_{\text{em}}$, describing the interaction between conduction electrons and the spin texture [9], are linked exactly once and again resemble the Hopf fibration [28].

Stability of Hopf solitons in chiral colloidal ferromagnets is enhanced by chirality (analogous to DMI) and typically strong perpendicular boundary conditions for $\boldsymbol{m}(\boldsymbol{r})$ at confining surfaces [18,19], with the latter setting the uniform far-field background $\boldsymbol{m}_0$. In solid-state ferromagnets the surface interactions of $\boldsymbol{m}(\boldsymbol{r})$ are weak and commonly neglected. However, strong effective PMA has been found at the interfaces of strained chiral magnets [32,33], magnetic metal and oxide interfaces [34], metallic multilayers [35], and chiral magnet-ferromagnet heterostructures [36,37]. The magnetic anisotropy energy in strained MnSi due to a lattice mismatch layer is $\sim$100 kJm$^{-3}$, which could enable an interfacial PMA within 0.1–1 mJm$^{-2}$ when induced in thin layers near the surface [33]. Experimentally measured PMA values of 1–2 mJm$^{-2}$ were reported for the interface of magnetic metal or alloy and oxide [34]. A multilayer structure FeGe/Fe/MgO can be designed such that the effect of PMA on the Fe/MgO interface is transferred to FeGe by exchange coupling [34,37]. These advances in controlling PMA bring about the possibility of using surface confinement and boundary conditions to control stability of solitonic $\boldsymbol{m}(\boldsymbol{r})$ structures, similar to the case of surface anchoring boundary conditions in liquid crystals [18,19]. Below we show that interfacial PMA stabilizes a host of solitonic structures, including hopfions (Fig. 1).

We perform energy-minimizing routines on a micromagnetic Hamiltonian of an isotropic chiral magnet that contains both bulk and surface terms [38]

$$E = \int_\Omega d^3\boldsymbol{r} \left[ \frac{\mathcal{J}_{\text{ex}}}{2}(\nabla \boldsymbol{m})^2 + \mathcal{D}_{\text{DMI}} \boldsymbol{m} \cdot (\nabla \times \boldsymbol{m}) - \mu_0 M_s m_z H \right] - \int_{\partial\Omega} d^2\boldsymbol{r} \frac{\mathcal{K}_s}{2}(\boldsymbol{m} \cdot \boldsymbol{n})^2 \quad (1)$$

where $\mathcal{J}_{\text{ex}}$ and $\mathcal{D}_{\text{DMI}}$ are Heisenberg exchange and DMI constants defining the helical wavelength $\lambda = 2\pi(\mathcal{J}_{\text{ex}}/\mathcal{D}_{\text{DMI}})$, $H$ and $M_s$ are the magnetic field applied along $\hat{z}$ and the saturated magnetization defining the Zeeman coupling energy, $\mathcal{K}_s$ characterizes the strength of PMA, $\boldsymbol{n}$ is the easy-axis direction for $\boldsymbol{m}(\boldsymbol{r})$ at the surface (chosen to be along the surface normal $\hat{z}$), $\Omega$ and $\partial\Omega$ are the magnet's volume and boundary, respectively. The strength of PMA can be quantified by an extrapolation length $\xi \equiv \mathcal{J}_{\text{ex}}/\mathcal{K}_s$, a virtual distance beyond the physical boundary where the hard boundary conditions are set, with $\xi = 0$ for infinitely strong PMA. To make our finding relevant to different material systems, we scale length in units of $\lambda$ and the magnetic field in units of $H_D \equiv \mathcal{D}_{\text{DMI}}^2/\mu_0 M_s \mathcal{J}_{\text{ex}}$, the critical field strength for field-polarized state in bulk chiral magnet [10]. Computer simulations are performed starting from an analytical ansatz [1] (previously also used to model hopfions in liquid crystals [18,19]) for a series of heterostructure geometries where chiral magnetic films of thickness $d \sim \lambda$ are confined between thin PMA-inducing layers (e.g., oxide or lattice mismatch layer) that define boundary conditions above and below the film, but not at its edges [Fig. 1(a)].

In nanodisks, ground-state hopfions with different Hopf indices (Fig. 2 and Supplemental Material [38], Fig. S1) arise from frustration that stems from competing terms in Eq. (1). The structural stability diagram also includes 2D Skyrmions and topologically trivial helical, modulated helical and conical states (Fig. 2), though these structures and their energetic costs are also altered by the boundary conditions (Supplemental Material [38], Fig. S1). Elementary $Q = 1$ hopfions are the ground state [Fig. 2(e)] at $\xi/\lambda \lesssim 0.05$, $H \lesssim 0.22 H_D$ and the diameter of the nanodisk $D \gtrsim 2.8\lambda$. Helical and 2D Skyrmion states are hindered by high surface energy costs and exist only at large $\xi$, whereas conical and field-polarized states appear at large fields and for tight lateral confinement. Hopfions with $Q = 2$ are stabile at $D \gtrsim 6\lambda$, and future studies can explore how geometry of nanostructures can predefine stability of hopfions with different $Q$. Supplemental Material [38], video S1 shows the structural evolution starting from a hopfion when the boundary conditions are removed, demonstrating the role of PMA in hopfion stability. A "half-hopfion" structure, a 3D analog of the 2D meron, can be stabilized for asymmetric boundary conditions [Figs. 2(f)-2(g)]. Free boundary conditions on the nanodisk edges result in the DMI-driven topologically trivial near-edge twist, consistent with the past studies of chiral magnetic nanostructures [43]. Computer-simulated Lorentz TEM images of a Hopf soliton for viewing directions along and perpendicular to $\hat{z}$ are shown in Fig. 2(d) differ from 2D Skyrmions and other solitonic structures, which may facilitate demonstration of hopfions in experiments.

Much like the Skyrmionic $A$ phase [44], hopfions can form a hexagonal 2D crystal in a film of thickness $d$ (Fig. 3). In the film geometry, the translationally invariant conical state becomes the ground state while the hopfion crystal is metastable with its metastability dependent on $H$, $\xi$ and $d$ (Fig. 3c). Hopfion stability is aided by strong boundary conditions at $d \sim \lambda$. At no fields, metastability conditions correspond to $d/\lambda \gtrsim 0.85$ and $\xi/\lambda < 0.056$, equivalent to $d \gtrsim 69$ nm and $\mathcal{K}_s > 1.6$ mJm$^{-2}$ for material parameters of FeGe and $d \gtrsim 15$ nm and $\mathcal{K}_s > 0.33$ mJm$^{-2}$ for MnSi (Supplemental Material [38], Table S1). Magnetic fields parallel to $\boldsymbol{m}_0$ effectively aid the confinement and lower the interfacial PMA required for stability. However, these fields also promote formation of conical states and a larger $d/\lambda$ is needed to gain stability by extra twisting. For example, hopfions can be metastable up to $H = 0.3 H_D$ at $d/\lambda = 1.2$, whereas magnetic fields antiparallel to $\boldsymbol{m}_0$ raise the needed interfacial PMA and lower $d/\lambda$. Lorentz TEM images of a hexagonal hopfion crystal [Fig. 3(f)] differs from the images of hexagonal Skyrmion crystals (see Refs. [11,45] and computer simulated images in Supplemental Material [38], Fig. S2 for comparison). Apart from the difference in pattern, the lattice constant is $\sim \lambda$ in a Skyrmion crystal [9, 46] and $\gtrsim 2.5\lambda$ in a hopfion crystal. Hopfions also emerge in the channel geometries that can be used in the racetrack memory [12] and other spintronics applications (Fig. 4). Lorentz TEM images and $\boldsymbol{m}(\boldsymbol{r})$ of these hopfions [Fig. 4(b)] qualitatively agree with the ones in films and nanodisks, though they are asymmetrically squeezed due to the lateral confinement only in one direction. Interestingly, the difference between the hopfion crystal metastable state and the corresponding stable state is often <1% of the equilibrium free energy. The effects due to magnetostatic energy and various types of bulk anisotropy on the stability of 2D hopfion crystals in thin films or 3D hopfion crystals require further investigations. Our findings call for a systematic study of various material parameters and confinement conditions under which such solitonic condensed matter phases could arise. Since hopfions of various $Q$ can help embedding localized twisted regions of $\boldsymbol{m}(\boldsymbol{r})$ in the uniform far-field ferromagnetic background, individual isolated hopfions could potentially arise during magnetic switching as transient or stable structures, though their stability in bulk materials remains an open question outside of the scope of present work.

Hopfions stabilized by fixed boundary conditions have their preimages closed and interlinked within the magnetic bulk [Fig. 1(b)]. The finite-strength interfacial PMA and the ensuing relaxed boundary condition allows the magnetization to deviate from $m_0$ at the surfaces. The largest deviation angle $\theta_c$ defines a subspace of points with polar angles $\theta < \theta_c$ on $\mathbb{S}^2$ that have partially "virtual" preimages closed outside $\Omega$ [Fig. 3(d) and 3(e)] but confined within the extended volume $\Omega \cup \Omega'$ defined by the extrapolation length. At $\xi = 0$, $\theta_c = 0$ and $\Omega' = 0$, but both increase with $\xi/\lambda$ until a threshold beyond which an abrupt transition to a structure without closed-loop preimages happens, making hopfions unstable. These hopfions can be analyzed by numerically integrating the Hopf index [19,47-49],

$$Q = \frac{1}{64\pi^2} \int_{\Omega \cup \Omega'} d^3\boldsymbol{r}\, \varepsilon^{ijk} A_i F_{jk}, \quad (2)$$

where $F_{ij} = \varepsilon_{abc} m^a \partial_i m^b \partial_j m^c$, $\varepsilon$ is the Levi-Civita totally antisymmetric tensor, $A_j$ is defined as $F_{ij} = (\partial_i A_j - \partial_j A_i)/2$, and the summation convention is assumed. For example, integration gives $Q = 0.9997 \approx 1$ at $d/\lambda = 0.85$ and $\xi/\lambda = 0.056$, consistent with $Q = 1$ obtained from the geometric analysis of preimage linking.

To conclude, through numerical modeling, we demonstrate ground-state and metastable hopfions in isotropic chiral magnets under nanoscale confinement of circular nanodisks, thin films, and channels, including metastable hexagonal hopfion crystals in a thin film. Further extension of our model to include magnetostatic energy and bulk anisotropy terms can alter the free energy landscape and could be leveraged to further enhance the stability of hopfions. The capability of encoding 1, 0, and −1 and other states in the topological charges of 3D Hopf solitons in a chiral magnet can lead to new architectures of data storage devices and other spintronics applications. Computer-simulated Lorentz TEM images of hopfions in common chiral ferromagnets like MnSi and FeGe sandwiched as nanostructures between layers inducing PMA exhibit unique features that will enable their experimental identification and potentially even assignment of Hopf index values.

We acknowledge discussions with P. Ackerman, M. Dennis, D. Foster, N. Nagaosa, H. Sohn, A. Thiaville, Y. Tokura, and X. Yu and funding from the U.S. Department of Energy, Office of Basic Energy Sciences, Division of Materials Sciences and Engineering, under Awards No. ER46921 and No. DE-SC0019293. This work utilized the RMACC Summit supercomputer, which is supported by the NSF (Grants No. ACI-1532235 and No. ACI-1532236), the University of Colorado Boulder and Colorado State University.

**Figures:**

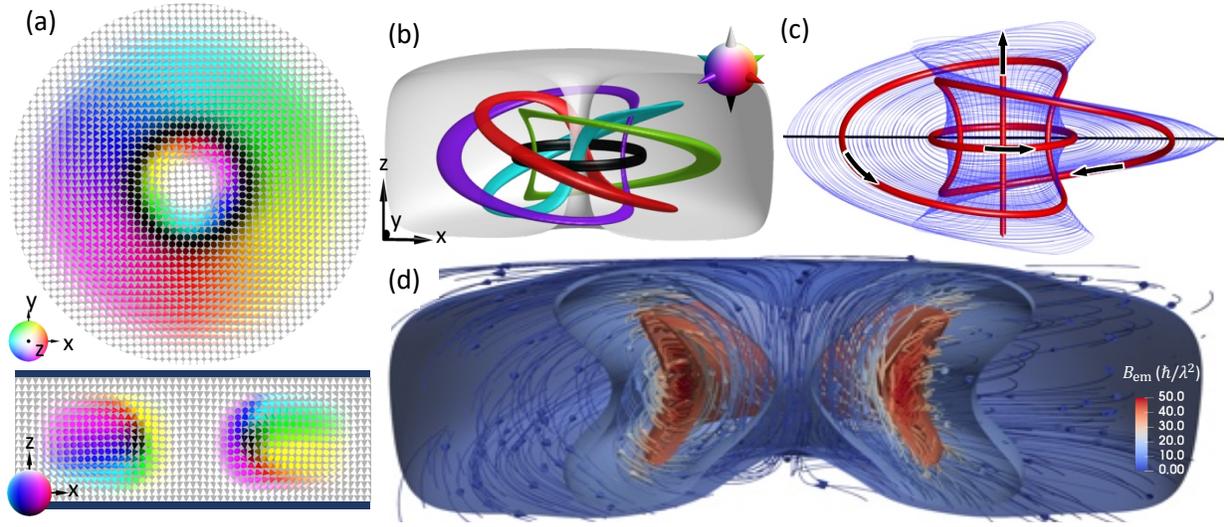

FIG. 1. 3D topological soltion – hopfion. (a) Mid-plane cross sections of a hopfion in the plane perpendicular to $\boldsymbol{m}_0$ (upper) and in the vertical plane containing $\boldsymbol{m}_0$ (lower). The magnetization fields are shown with cones colored according to the corresponding points on $\mathbb{S}^2$ (lower-left insets). In the x-z cross section, the black stripes at the top and bottom indicate fixed boundary conditions that can be achieved, for example, using thin films of a different material (e.g., oxide or lattice mismatch layer). (b) The 3D preimages of points on $\mathbb{S}^2$ indicated as cones in the upper-right inset. The linking number of preimages yields $Q = 1$. (c) Streamlines of $\boldsymbol{B}_{\text{em}}$ form the Hopf fibration. Subsets of streamlines originating from points on a horizontal black line are illustrated by the blue lines, with some highlighted by red tubes to show interlinking of the ensuing closed loops, with the linking number 1. (d) Visualization of $\boldsymbol{B}_{\text{em}}$ by the isosurfaces of constant magnitude and streamlines with cones indicating directions.

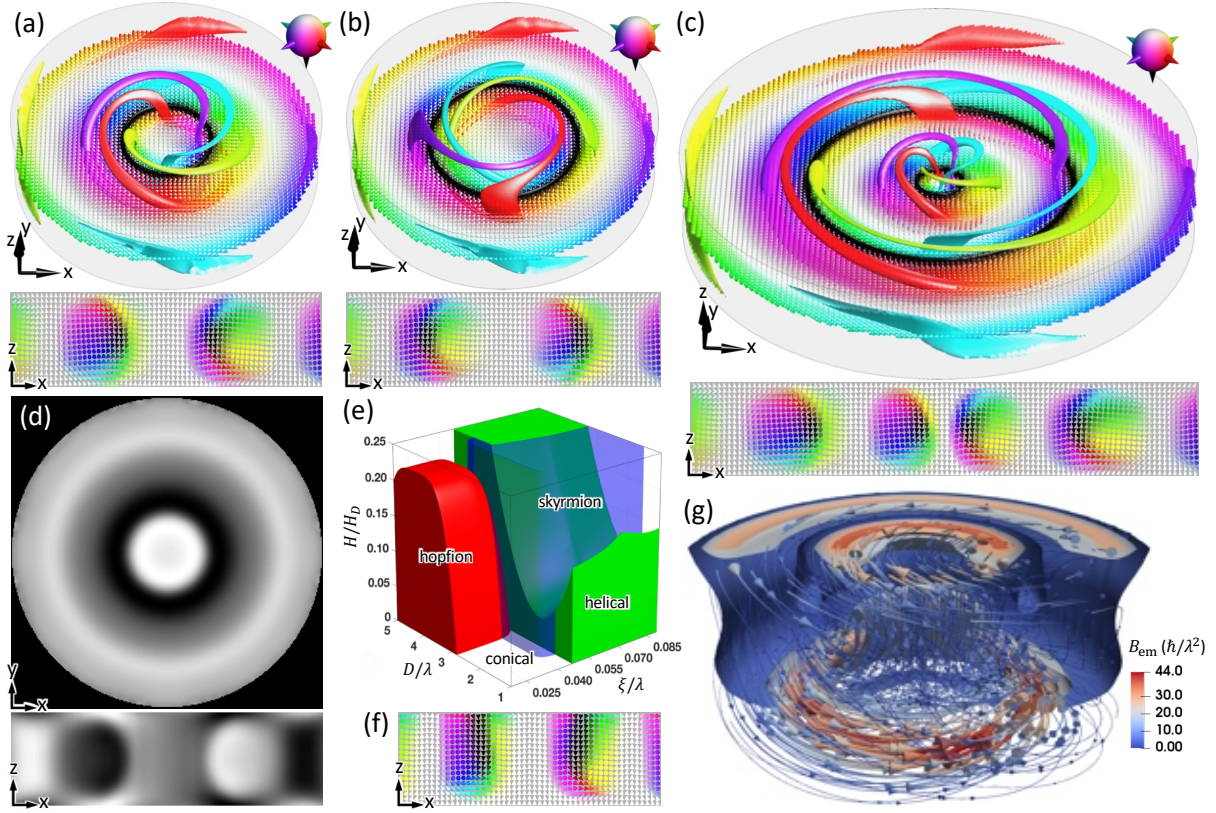

FIG. 2. Stable hopfions in circular nanodisks. (a),(b),(c) Horizontal midplane cross sections (upper) and vertical midplane cross sections (lower) of hopfions with $Q = 1, -1$ and 2, respectively, shown along with preimages of points on $\mathbb{S}^2$ (corresponding to cones in the upper-right insets). (d) Computer-simulated Lorentz TEM images of a $Q = 1$ hopfion shown in (a) for viewing directions along $\hat{z}$ (upper) and perpendicular to it (lower). (e) Ground-state stability diagram of solitonic structures in nanodisks. The parameter space of stable hopfions, Skyrmions and helical states are shown in red, blue and green, respectively, and that for the conical state is left blank. (f) A half-hopfion with PMA only on the bottom interface for $d = 0.9\lambda$ and $D = 4\lambda$. (g) Visualization of half-hopfion's $\boldsymbol{B}_{em}$ derived from (f) by the isosurfaces colored by magnitude and streamlines with cones indicating directions.

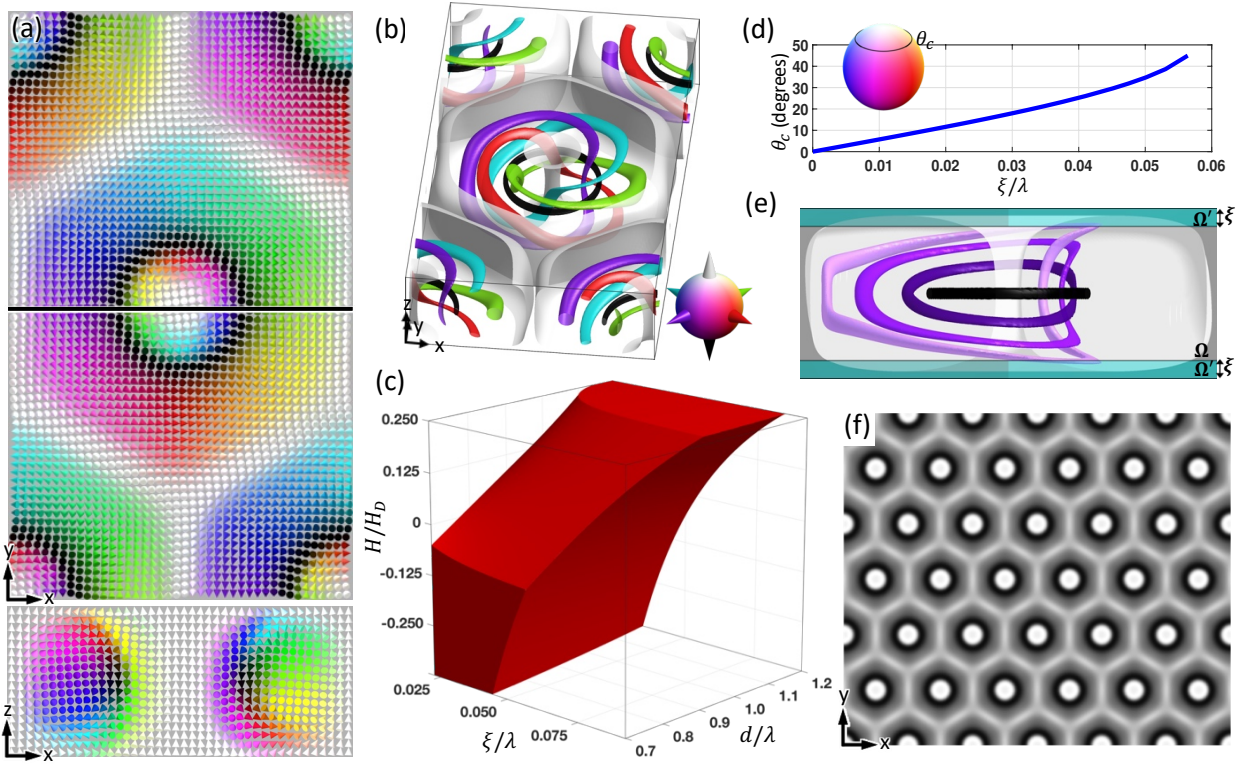

FIG. 3. Hexagonal hopfion crystal in a thin film of a chiral magnet. (a) Midplane cross sections of a unit cell of the hopfion crystal in a plane perpendicular to $\boldsymbol{m}_0$ (upper) and in the vertical plane (indicated by a black line in the upper panel) containing $\boldsymbol{m}_0$ (lower). (b) 3D preimages of points on $\mathbb{S}^2$ indicated by cones in the lower-right inset. (c) Diagram of metastability of a hopfion crystal (shown in red) vs. $\xi/\lambda$, $d/\lambda$ and $H/H_D$. (d) Dependence of $\theta_c$ on $\xi/\lambda$ at $d/\lambda = 0.9$. Shown in the inset is the target $\mathbb{S}^2$ with the boundary at $\theta = 45°$ for $\xi/\lambda = 0.056$. (e) A unit cell of the hopfion crystal confined in the space $\Omega \cup \Omega'$ extended by $\xi$. (f) A Lorentz TEM image of a 2D hopfion crystal.

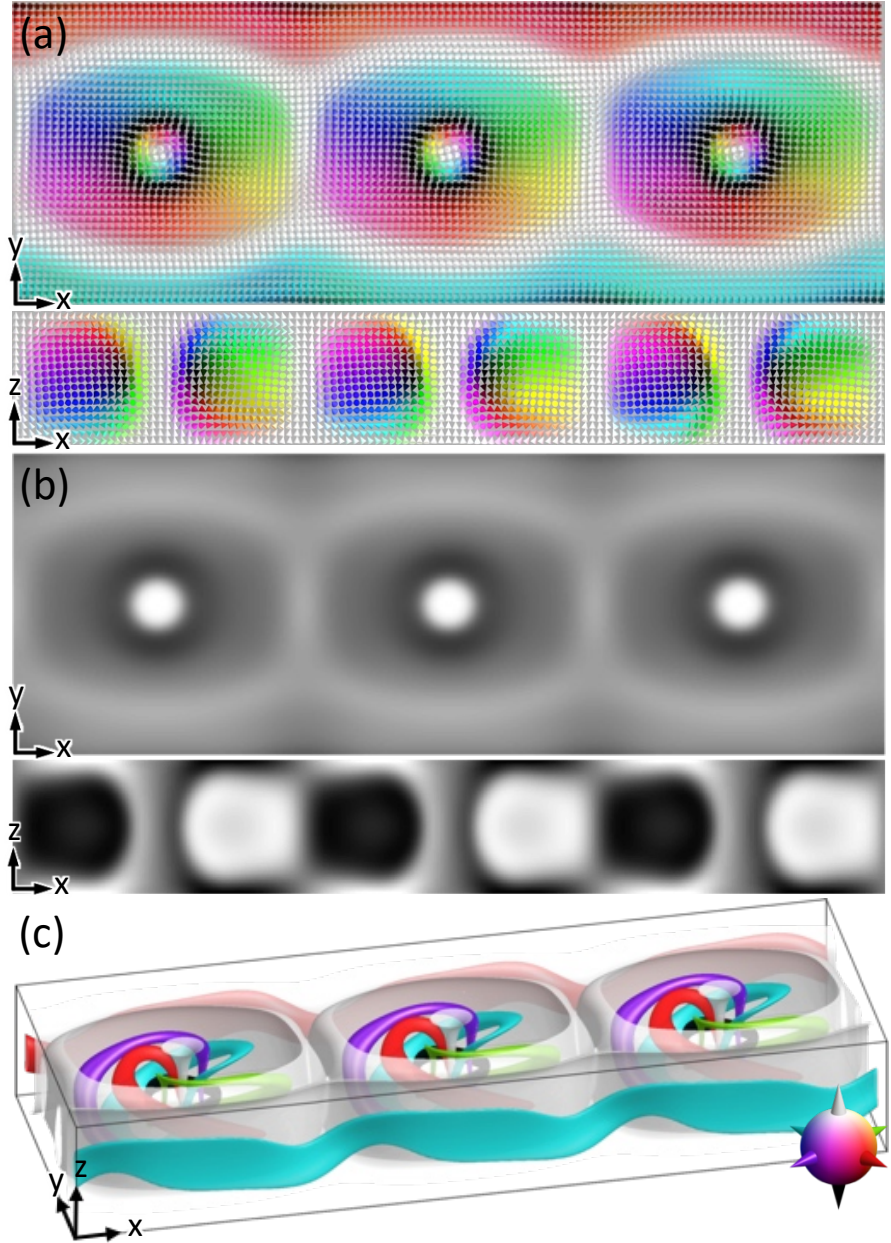

FIG. 4. Hopfions in magnetic channels shown for three unit cells. (a) Midplane cross sections of a channel of hopfions in the plane perpendicular to $\boldsymbol{m}_0$ (upper) and in the vertical plane containing $\boldsymbol{m}_0$ (lower). (b) Computer-simulated Lorentz TEM images of hopfions in a channel when viewed along $\hat{z}$ (upper) and orthogonally to it (lower). (c) 3D preimages of points on $\mathbb{S}^2$ indicated as cones in the lower-right inset.

**Supplemental Material for "Static Hopf solitons and knotted emergent fields in solid-state noncentrosymmetric magnetic nanostructures"**

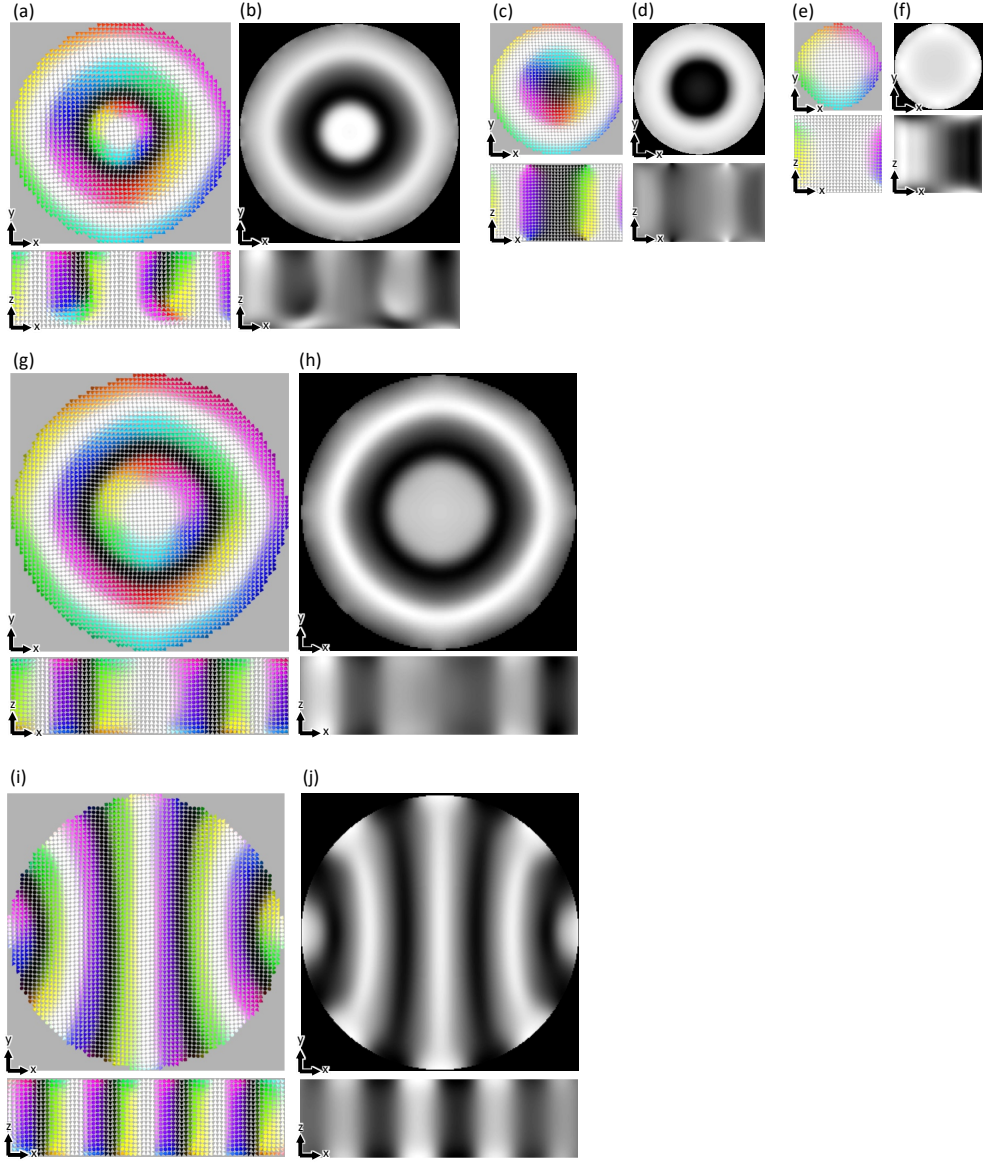

FIG. S1. Magnetization structures stabilized in circular nanodisks with $d = \lambda$ of different diameters and boundary conditions (additional details for Fig. 2). (a, c, e, g and i) Mid-plane cross-sections of structures at no applied field in the plane perpendicular to $\boldsymbol{m}_0$ (upper) and in the vertical plane containing $\boldsymbol{m}_0$ (lower). (a) A half-hopfion obtained at $D = 3.2\lambda$ and $\xi/\lambda = 0.023$ on the bottom interface and no PMA on the top interface (c) A skyrmion obtained at $D = 2\lambda$, $\xi/\lambda = 0.045$ on both interfaces. (e) An (axially-symmetric) conical state obtained at $D = 1.33\lambda$ and $\xi/\lambda = 0.045$ on both interfaces. (g) A modulated helical state (skyrmion-antiskyrmion pair) that appears as a result of a hopfion transformation upon switching PMA off at $D = 4\lambda$. (i) A helical state obtained at $D = 4\lambda$ and no PMA on either interface. (b, d, f, h and j) Computer-simulated Lorentz TEM images of structures corresponding to (a, c, e, g and i), respectively, when viewed along $\hat{z}$ (upper) and along $\hat{y}$ (lower).

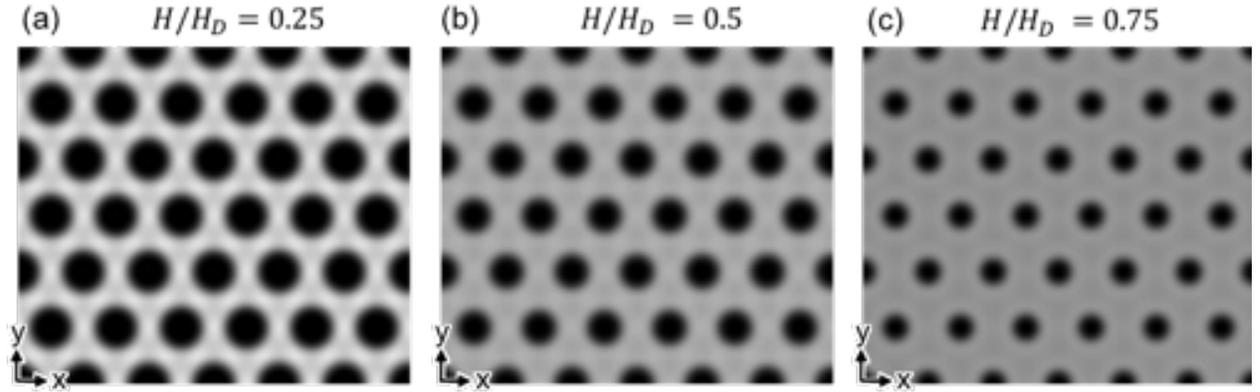

FIG. S2. Computer-simulated Lorentz TEM images of a 2D skyrmion crystal at different applied magnetic fields.

Table S1. Material parameters of FeGe at 200K [1] and MnSi [2].

| Material | $\mathcal{J}_{ex}$ | $\mathcal{D}_{DMI}$ | $\lambda = 2\pi(\frac{\mathcal{J}_{ex}}{\mathcal{D}_{DMI}})$ | $\mu_0 M_s$ |
|---|---|---|---|---|
| FeGe | 7.290 pJm$^{-1}$ | 0.567 mJm$^{-2}$ | 80.8 nm | 0.14 T |
| MnSi | 0.32 pJm$^{-1}$ | 0.115 mJm$^{-2}$ | 17.5 nm | 0.19 T |

**Supplementary Methods**

In the continuum approximation, the effective Hamiltonian of an isotropic bulk chiral magnet is given by the volume integral in Eq. (S1)

$$E = \int_\Omega d^3r\, h = \int_\Omega d^3r \left[\frac{\mathcal{J}_{ex}}{2}(\nabla \boldsymbol{m})^2 + \mathcal{D}_{DMI}\boldsymbol{m}\cdot(\nabla \times \boldsymbol{m}) - \mu_0 M_s \boldsymbol{m}\cdot \boldsymbol{H}\right] \quad (S1)$$

where $\mathcal{J}_{ex}$ and $\mathcal{D}_{DMI}$ describe the magnitude of Heisenberg exchange and the Dzyaloshinskii-Moriya interaction (DMI) and the resulting helical wavelength is $\lambda = 2\pi(\mathcal{J}_{ex}/\mathcal{D}_{DMI})$. The last term in Eq. (S1) is the Zeeman energy density describing the coupling between the magnetization and the applied magnetic field $\boldsymbol{H}$, where $M_s$ is the saturated magnetization. When taking into account magnetic anisotropy energy on the boundaries, Eq. (S1) is supplemented with a surface term

$$-\int_{\partial\Omega} d^2r\, \frac{\mathcal{K}_s}{2}(\boldsymbol{m}\cdot\boldsymbol{n})^2 \quad (S2)$$

where $\mathcal{K}_s$ characterizes the strength of the interfacial magnetic anisotropy and $\boldsymbol{n}$ is the easy-axis direction for $\boldsymbol{m}(\boldsymbol{r})$.

For the static field configurations to be observed in the magnetic system, they need to emerge as local or global minima of the Hamiltonian given by Eqs. (S1) and (S2). Numerical modeling of the energy minimization of $\boldsymbol{m}(\boldsymbol{r})$ is performed by a variational-method-based relaxation routine which also includes the boundary effects [3-5]. After discretizing the computational volume, each node is identified to be either a bulk node or a boundary node. At each iteration of the numerical simulation, $\boldsymbol{m}(\boldsymbol{r})$ is updated based on an update formula derived from the Lagrange equation of the system,

$$m_i^{new} = m_i^{old} - \frac{\text{MSTS}}{2}[E]_{m_i} \quad (S3)$$

where the subscript $i$ denotes spatial coordinates, $[E]_{m_i}$ denotes the functional derivative of $E$ with respect to $m_i$. To include the boundary effects, $[E]_{m_i}$ at a boundary node takes a different form than at a bulk node. Specifically

$$[E]_{m_i} = \frac{\partial h}{\partial m_i} - \nabla \cdot \frac{\partial h}{\partial \nabla m_i} \text{ (bulk) (S4)}$$

$$[E]_{m_i} = \frac{\partial h}{\partial \nabla m_i} \cdot \hat{\boldsymbol{e}} - \mathcal{K}_s (\boldsymbol{m} \cdot \boldsymbol{n}) n_i \text{ (boundary) (S5)}$$

where $\hat{\boldsymbol{e}}$ is the surface normal on the boundary. MSTS is the maximum stable time step in the minimization routine, determined by the values of material parameters and the spacing of the computational grid [3-5]. Since the above update formulae do not guarantee unit modulus of $\boldsymbol{m}(\boldsymbol{r})$, $\boldsymbol{m}(\boldsymbol{r})$ is normalized after each iteration. The steady-state stopping condition is determined by monitoring the change in the spatially averaged functional derivatives over iterations. When this value approaches zero, the system is implied to be in a state corresponding to the energy minimum, and the relaxation routine is terminated. By iteratively solving the Euler-Lagrange equation derived from Eqs. (S1) and (S2), energy minima and evolution dynamics of the magnetization $\boldsymbol{m}(\boldsymbol{r})$ can be found.

The 3D spatial discretization is performed on large 3D square-periodic grids with 24 to 72 grid points per helical wavelength $\lambda$. The spatial derivatives are calculated by finite difference methods with second-order accuracies based on central difference for the bulk nodes and one-sided differences for the boundary nodes, respectively, allowing us to minimize discretization-related artifacts. For simulating hexagonal hopfion crystals, periodic boundary conditions are imposed in both lateral directions. Periodic boundary conditions are imposed along one direction in the case of hopfions in channels. Both the analytical ansatz configurations [6,7] and hopfion structures in liquid crystals [3,4] are used as initial conditions.

To construct a preimage of a point on $\mathbb{S}^2$ within the 3D volume of the static topological solitons, we calculate a scalar field defined as the difference between the magnetization field $\boldsymbol{m}(\boldsymbol{r})$ and a unit vector defined by the target point on $\mathbb{S}^2$. The preimage is then visualized with the help of the isosurfaces of a small value in this ensuing scalar field. The Hopf index $Q$ of a 3D topological soliton can be geometrically interpreted as the linking number of each pair of distinct preimages [6]. By choosing the circulation of the preimage of the north pole on $\mathbb{S}^2$ to be along $\boldsymbol{m}_0$ through the center of the topological solitons, the circulations of all other preimages are determined by smoothly moving away from the north pole and exploring $\mathbb{S}^2$. Note this choice does not affect the resulting linking number. The linking number of preimages is then defined as half the total number of crossings, with the sign of each crossing defined by the convention based on the right-hand rule [6,8]. Within this procedure, by flattening the right hand, we extend the fingers in the direction along the circulation of one preimage with the palm facing the other. The sign of the crossing is then positive if the circulation of the other preimage and the thumb's direction point toward the same side with respect to the first preimage, and negative otherwise. The values of $Q$ determined via this approach is consistent with the one obtained via numerical integration, as described below.

To numerically calculate the Hopf index by integrating Eq. (2) in the main text, we can define $b^i \equiv \varepsilon^{ijk} F_{jk}$ such that $b^i = \varepsilon^{ijk}(\partial_j A_k - \partial_k A_j)/2 = \varepsilon^{ijk}\partial_j A_k$ and $A$ can be understood as the vector potential of the vector field $\boldsymbol{b}$, and $Q$ can be rewritten as $Q = 1/64\pi^2 \int d^3\boldsymbol{r}\, \boldsymbol{b} \cdot \boldsymbol{A}$. After calculating $\boldsymbol{b}$ from $\boldsymbol{m}(\boldsymbol{r})$, the vector potential $\boldsymbol{A}$ can be obtained by numerically integrating $\boldsymbol{b}$ [8]. The numerically integrated $Q$ approaches that determined geometrically by the linking number of preimages as the solitonic field configurations are interpolated on a finer grid. In this work, the fields were interpolated on a grid 8 times finer than the original grid, for which the free-energy minimization was performed.

To simulate Fresnel mode images of Lorentz transmission electron microscopy, we make use of the following equation, which describes the intensity of an image at small defocus $\Delta$ for a thin film normal to the electron beam direction $\hat{l}$ [9]

$$I(\boldsymbol{r}) = 1 - \int_0^d dl\, \Delta \frac{e\mu_0 \lambda_e}{2\pi\hbar} \big(\nabla \times \boldsymbol{m}(\boldsymbol{r})\big) \cdot \hat{\boldsymbol{l}} \quad \text{(S6)}$$

where $\lambda_e$ is the electron wavelength, $d$ is the film thickness, and $e$, $\mu_0$ and $\hbar$ are the electron charge, vacuum permeability, and the reduced Planck constant, respectively. The contrast of each image is then normalized.

**Supplementary References:**